# A Public Key Cryptoscheme Using Bit-pairs and Probabilistic Mazes [*]

Shenghui Su [1, 4 (✉)], Shuwang Lü [2, 5], Maozhi Xu [3], and Tao Xie [4]

[1] School of Computers, Nanjing University of Science and Technology, Nanjing 210094, PRC
[2] School of Computers, University of Chinese Academy of Sciences, Beijing 100039, PRC
[3] School of Mathematics, Peking University, Beijing 100871, PRC
[4] School of Computers, National University of Defense Technology, Changsha 410073, PRC
[5] Laboratory of Computational Complexity, BFID Corporation, Beijing 100098, PRC

**Abstract.** This paper gives the definition and property of a bit-pair shadow, and devises the three algorithms of a public key cryptoscheme called JUOAN that is based on a multivariate permutation problem and an anomalous subset product problem to which no subexponential time solutions are found so far, and regards a bit-pair as a manipulation unit. The authors demonstrate that the decryption algorithm is correct, deduce the probability that a plaintext solution is nonunique is nearly zero, analyze the security of the new cryptoscheme against extracting a private key from a public key and recovering a plaintext from a ciphertext on the assumption that an integer factorization problem, a discrete logarithm problem, and a low-density subset sum problem can be solved efficiently, and prove that the new cryptoscheme using random padding and random permutation is semantically secure. The analysis shows that the bit-pair method increases the density $D$ of a related knapsack to a number more than 1, and decreases the modulus length $\lceil \lg M \rceil$ of the new cryptoscheme to 464, 544, or 640.

**Keywords:** Public key cryptoscheme; Semantical security; Bit-pair shadow; Random padding; Anomalous subset sum problem; Compact sequence

## 1 Introduction

There is a sensitive relation between a subset sum problem and a shortest vector problem [1][2].

Let $\{u_1, \ldots, u_n\}$ be a positive integer sequence, namely a knapsack, and $\mathbb{L}$ be a lattice spanned by $n + 1$ vectors which compose a lattice basis. Then, each of the prior $n$ vectors of the lattice basis is relevant to $u_i$ respectively, where $i = 1, 2, \ldots,$ or $n$, and the last vector is relevant to a subset sum, namely the element sum of a subset of $\{u_1, \ldots, u_n\}$ [1][2].

If the density of the knapsack is less than 0.9408, the shortest vector will occur in the reduced basis with high probability which indicates it is very likely that the subset sum problem will have a solution in polynomial time. If the density of the knapsack is greater than 1, there will exist many vectors which have the same shortest length in $\mathbb{L}$ which indicates that the special shortest vector as the solution to the subset sum problem will not occur in the reduced basis with high probability; or the solution to the subset sum problem is not among the shortest vectors [2].

In [3], we bring forward a prototypal public key cryptosystem called REESSE1+ which is based on the three new provable problems, contains the five algorithms, and is utilized for data encryption and digital signing.

In REESSE1+, a ciphertext is defined as $\bar{G} \equiv \prod_{i=1}^{n} C_i^{b_i}$ (% $M$), an anomalous subset product problem (ASPP), where $M$ is a prime modulus, $b_i$ is the bit shadow of a bit $b_i$, $\{C_i\}$ is a public key sequence, and $n$ is the bit-length of a plaintext block [3].

Let $C_1 \equiv g^{u_1}$ (% $M$), ..., $C_n \equiv g^{u_n}$ (% $M$), and $\bar{G} \equiv g^v$ (% $M$), where $g$ is a generator of $(\mathbb{Z}_{M}^{*}, \cdot)$ which can be found when the modulus $M < 2^{1024}$ is factorized in tolerable sub-exponential time [4]. Then solving $\bar{G} \equiv \prod_{i=1}^{n} C_i^{b_i}$ (% $M$) for $b_1 \ldots b_n$ is equivalent to solving

$$b_1 u_1 + \ldots + b_n u_n \equiv v \ (\%\ \bar{M}). \tag{1}$$

where $v$ may be substituted with $v + k\bar{M}$ along with $k \in [0, n-1]$ [5].

---

[*] This work is supported by MOST with Project 2009AA01Z441 and NSFC with Project 61472476. Email: reesse@126.com.
Referring to: Theoretical Computer Science, v654, Nov 2016, pp.113–127.

Equation (1) is called an anomalous subset sum problem (ASSP) due to every $b_i \in [0, n]$ [3].

Likewise due to every $b_i \in [0, n]$, $\{u_1, ..., u_n\}$ is called a compact sequence [6].

May convert an ASSP into a subset sum problem (SSP) through converting $u_i$ to a binary number, and thus the density of an ASSP knapsack is defined as

$$D = \sum_{i=1}^{n} \lceil \lg n \rceil / \lceil \lg M \rceil = n \lceil \lg n \rceil / \lceil \lg M \rceil. \quad (2)$$

Transparently, the parameters $\lceil \lg M \rceil$ and $n$ have an important influence on the value of $D$.

In REESSE1+, there is $D < 1$, which means that the original solution to an ASSP may possibly be found through the LLL lattice basis reduction algorithm [7][8]. The LLL reduction algorithm is famous for it has a fatal threat to the classical MH knapsack cryptosystem [9] which accepts a public key and a plaintext block, and produces a ciphertext in the form of a subset sum.

To avoid the low density of a knapsack from an ASPP and to decrease the modulus length of a cryptosystem, on the basis of REESSE1+, we propose a new cryptoscheme called JUOAN which treats a bit-pair as an operation unit, brings randomicity into a ciphertext when a bit string is encrypted, and is proved to be semantically secure.

Throughout this paper, unless otherwise specified, $n \geq 80$ is the bit-length of a plaintext block, $\tilde{n} = n + 16$ is the bit-length of a padded plaintext block, the sign % denotes "modulo", $\bar{M}$ does "$M - 1$" with $M$ prime, $\lg x$ means the logarithm of $x$ to the base 2, $\neg$ does the opposite value of a bit, $Ᵽ$ does the maximal prime allowed in coprime sequences, $|x|$ does the absolute value of a number $x$, $\|x\|$ does the order of an element $x \% M$, $|S|$ does the size of a set $S$, $\gcd(a, b)$ represents the greatest common divisor of two integers, and "bos" indicates the number of bit operations. Without ambiguity, "$\% M$" is usually omitted in expressions.

## 2 Several Definitions

### 2.1 A Coprime Sequence

***Definition 1:*** If $A_1, ..., A_n$ are $n$ pairwise distinct positive integers such that $\forall A_i, A_j (i \neq j)$, either $\gcd(A_i, A_j) = 1$ or $\gcd(A_i, A_j) = F \neq 1$ with $(A_i / F) \nmid A_k$ and $(A_j / F) \nmid A_k \; \forall k \neq i, j \in [1, n]$, these integers are called a coprime sequence, denoted by $\{A_1, ..., A_n\}$, shortly $\{A_i\}$.

Note that the elements of a coprime sequence are not necessarily pairwise coprime, but a sequence of which all the elements are pairwise coprime is a coprime sequence.

For example, {15, 61, 163, 31, 37, 509, 21, 1669}, {37, 23, 7, 1009, 3, 1999, 937, 29}, {3607, 17, 59, 97, 1021, 211, 863, 2039}, and {10, 211, 127, 307, 14, 1021, 2017, 263} are four coprime sequences separately.

***Property 1:*** Let $\{A_1, ..., A_n\}$ be a coprime sequence. If randomly select $k \in [1, n]$ elements $A_{x_1}, ..., A_{x_k}$ from the sequence, then the mapping from a subset $\{A_{x_1}, ..., A_{x_k}\}$ to a subset product $G = \prod_{i=1}^{k} A_{x_i}$ is one-to-one, namely the mapping from $b_1...b_n$ to $G = \prod_{i=1}^{n} A_i^{b_i}$ is one-to-one, where $b_1...b_n$ is a bit string.

Refer to [3] for its proof.

### 2.2 A Bit Shadow

***Definition 2:*** Let $b_1...b_n \neq 0$ be a bit string. Then $ƀ_i$ with $i \in [1, n]$ is called a bit shadow if it comes from such a rule:

① $ƀ_i = 0$ if $b_i = 0$,

② $ƀ_i = 1 +$ the number of successive 0-bits before $b_i$ if $b_i = 1$, or

③ $ƀ_i = 1 +$ the number of successive 0-bits before $b_i +$ the number of successive 0-bits after the rightmost 1-bit if $b_i$ is the leftmost 1-bit.

***Fact 1:*** Let $ƀ_1...ƀ_n$ be the bit shadow string of $b_1...b_n \neq 0$. Then there is $\sum_{i=1}^{n} ƀ_i = n$.

*Proof:*

According to Definition 2, every bit of $b_1...b_n$ is considered into $\sum_{i=1}^{k} ƀ_{x_i}$, where $k \leq n$, and $ƀ_{x_1}, ..., ƀ_{x_k}$ are 1-bit shadows in the string $ƀ_1...ƀ_n$, and thus there is $\sum_{i=1}^{k} ƀ_{x_i} = n$.

On the other hand, there is $\sum_{j=1}^{n-k} ƀ_{y_j} = 0$, where $ƀ_{y_1}, ..., ƀ_{y_{n-k}}$ are 0-bit shadows.

In total, there is $\sum_{i=1}^{n} \flat_i = n$. □

**Property 2:** Let $\{A_1, \ldots, A_n\}$ be a coprime sequence, and $\flat_1 \ldots \flat_n$ be the bit shadow string of $b_1 \ldots b_n \neq 0$. Then the mapping from $b_1 \ldots b_n$ to $G = \prod_{i=1}^{n} A_i^{\flat_i}$ is one-to-one.

*Proof:*

Let $b_1 \ldots b_n$ and $b'_1 \ldots b'_n$ be two different nonzero bit strings, and $\flat_1 \ldots \flat_n$ and $\flat'_1 \ldots \flat'_n$ be the two corresponding bit shadow strings.

① If $\flat_1 \ldots \flat_n = \flat'_1 \ldots \flat'_n$, then by Definition 2, there is $b_1 \ldots b_n = b'_1 \ldots b'_n$.

Again, for any arbitrary bit shadow string $\flat_1 \ldots \flat_n$, there always exists a preimage $b_1 \ldots b_n$, namely the mapping from $b_1 \ldots b_n$ to $\flat_1 \ldots \flat_n$ is surjective.

Thus, the mapping from $b_1 \ldots b_n$ to $\flat_1 \ldots \flat_n$ is one-to-one.

② Obviously the mapping from $\flat_1 \ldots \flat_n$ to $\prod_{i=1}^{n} A_i^{\flat_i}$ is surjective.

Again, presuppose that $\prod_{i=1}^{n} A_i^{\flat_i} = \prod_{i=1}^{n} A_i^{\flat'_i}$ for $\flat_1 \ldots \flat_n \neq \flat'_1 \ldots \flat'_n$.

Since $\{A_1, \ldots, A_n\}$ is a coprime sequence, and $A_i^{\flat_i}$ either equals 1 with $\flat_i = 0$ or contains the same prime factors as those of $A_i$ with $\flat_i \neq 0$, we can obtain $\flat_1 \ldots \flat_n = \flat'_1 \ldots \flat'_n$ from $\prod_{i=1}^{n} A_i^{\flat_i} = \prod_{i=1}^{n} A_i^{\flat'_i}$. It is in direct contradiction to $\flat_1 \ldots \flat_n \neq \flat'_1 \ldots \flat'_n$, which indicates that the mapping from $\flat_1 \ldots \flat_n$ to $\prod_{i=1}^{n} A_i^{\flat_i}$ is injective [10].

Thus, the mapping from $\flat_1 \ldots \flat_n$ to $\prod_{i=1}^{n} A_i^{\flat_i}$ is one-to-one.

By transitivity, the mapping from $b_1 \ldots b_n$ to $\prod_{i=1}^{n} A_i^{\flat_i}$ is also one-to-one. □

## 2.3 A Bit-pair Shadow

To make the modulus $M$ of the new cryptoscheme comparatively small, we will utilize the idea of a bit-pair string with 2 bits to 3 items.

In this wise, the length of a coprime sequence is changed to $3n/2$, namely $\{A_1, \ldots, A_n\}$ is substituted with $\{A_1, A_2, A_3, \ldots, A_{3n/2-2}, A_{3n/2-1}, A_{3n/2}\}$ that may be logically orderly partitioned into $n/2$ triples of which each comprises 3 elements: $A_{3j-2}, A_{3j-1}, A_{3j}$ with $j \in [1, n/2]$. Likewise, a non-coprime sequence $\{C_1, \ldots, C_n\}$ is substituted with $\{C_1, C_2, C_3, \ldots, C_{3n/2-2}, C_{3n/2-1}, C_{3n/2}\}$, where $(C_{3j-2}, C_{3j-1}, C_{3j})$ with $j \in [1, n/2]$ is acquired from $(A_{3j-2}, A_{3j-1}, A_{3j})$ and other private parameters.

**Definition 3:** Let $\{A_{3j-2}, A_{3j-1}, A_{3j} \mid j = 1, \ldots, n/2\}$ be a coprime sequence. Orderly partition a bit string $b_1 \ldots b_n$ into $n/2$ pairs $B_1, \ldots, B_{n/2}$, where $B_j$ with $j \in [1, n/2]$ has four states: 00, 01, 10, and 11 which correspond to 1, $A_{3j-2}, A_{3j-1}$, and $A_{3j}$ respectively. Then $B_1, \ldots, B_{n/2}$ is called a bit-pair string, shortly $B_1 \ldots B_{n/2}$.

**Property 3:** Let $\{A_{3j-2}, A_{3j-1}, A_{3j} \mid j=1, \ldots, n/2\}$ be a coprime sequence, and $B_1 \ldots B_{n/2}$ be a nonzero bit-pair string. Then the mapping from $B_1 \ldots B_{n/2}$ to $G' = \prod_{i=1}^{n/2} (A_{3(i-1)+B_i})^{\lceil B_i/3 \rceil}$ with $A_0 = 1$ is one-to-one, where $\lceil B_i/3 \rceil = 0$ or 1, and $G'$ is called a coprime subsequence product.

Its proof is parallel to that of Property 1 in [3].

**Definition 4:** Let $B_1 \ldots B_{n/2}$ be a nonzero bit-pair string. Then $\flat_i$ with $i \in [1, n/2]$ is called a bit-pair shadow if it comes from such a rule:

① $\flat_i = 0$ if $B_i = 00$,

② $\flat_i = 1 +$ the number of successive 00-pairs before $B_i$ if $B_i \neq 00$, or

③ $\flat_i = 1 +$ the number of successive 00-pairs before $B_i +$ the number of successive 00-pairs after the rightmost non-00-pair if $B_i$ is the leftmost non-00-pair.

For example, let $B_1 \ldots B_4 = 10010100$ or $00100011$, and then $\flat_1 \ldots \flat_4 = 2110$ or $0202$.

**Fact 2:** Let $\flat_1 \ldots \flat_{n/2}$ be the bit-pair shadow string of $B_1 \ldots B_{n/2} \neq 0$. Then there is $\sum_{i=1}^{n/2} \flat_i = n/2$.

*Proof:*

According to Definition 4, every pair of $\flat_1 \ldots \flat_{n/2}$ is considered into $\sum_{i=1}^{k} \flat_{x_i}$, where $k \leq n/2$, and $\flat_{x_1}, \ldots, \flat_{x_k}$ are non-00-pair shadows in the string $\flat_1 \ldots \flat_{n/2}$, and thus there is $\sum_{i=1}^{k} \flat_{x_i} = n/2$.

On the other hand, there is $\sum_{j=1}^{n/2-k} \flat_{y_j} = 0$, where $\flat_{y_1}, \ldots, \flat_{y_{n/2-k}}$ are 00-pair shadows.

In total, there is $\sum_{i=1}^{n/2} \flat_i = n/2$. □

**Property 4**: Let $\{A_{3j-2}, A_{3j-1}, A_{3j} \mid j = 1, \ldots, n/2\}$ be a coprime sequence, and $\mathcal{B}_1\ldots \mathcal{B}_{n/2}$ be the bit-pair shadow string of $B_1\ldots B_{n/2} \neq 0$. Then the mapping from $B_1\ldots B_{n/2}$ to $G = \prod_{i=1}^{n/2} (A_{3(i-1)+B_i})^{\mathcal{B}_i}$ with $A_0 = 1$ is one-to-one, where $G$ is called an anomalous coprime subsequence product.

Its proof is parallel to that of Property 2 in Section 2.2.

## 2.4 A Lever Function

**Definition 5**: The secret parameter $\ell(i)$ in the key transform of a public key cryptoscheme is called a lever function, if it has the following features:

- $\ell(.)$ is an injection from the domain $\{1, \ldots, n\}$ to the codomain $\Omega \subset \{5, \ldots, \overline{M}\}$ with $\overline{M}$ large;
- the mapping between $i$ and $\ell(i)$ is established randomly without an analytical expression;
- an attacker has to be faced with all the arrangements of $n$ elements in $\Omega$ when extracting a related private key from a public key;
- the owner of a private key only needs to consider the accumulative sum of $n$ elements in $\Omega$ when recovering a related plaintext from a ciphertext.

The latter two points manifest that if $n$ is large enough, it is infeasible for the attacker to search all the permutations of elements in $\Omega$ exhaustively while the decryption of a normal ciphertext is feasible in polynomial time in $n$. Thus, there are the large amount of calculation on $\ell(.)$ at "a public terminal", and the small amount of calculation on $\ell(.)$ at "a private terminal".

Notice that ① in arithmetic modulo $\overline{M}$, $-x$ represents $\overline{M} - x$; ② considering the speed of decryption, the absolute values of all the elements should be comparatively small; ③ the lower limit 5 will make seeking the root $W$ from $W^{\ell(i)} \equiv A_i^{-1} C_i$ (% $M$) face an unsolvable Galois group when the value of $A_i \leq 1201$ is guessed [11].

**Property 5 (Indeterminacy of $\ell(.)$)**: Let $\delta = 1$ and $C_i \equiv (A_i W^{\ell(i)})^\delta$ (% $M$) with $\ell(i) \in \Omega = \{5, 6, \ldots, n+4\}$ and $A_i \in \Lambda = \{2, 3, \ldots, \mathcal{P} \mid \mathcal{P} \leq 1201\}$ for $i = 1, \ldots, n$. Then $\forall W (\|W\| \neq \overline{M}) \in (1, \overline{M})$ and $\forall x, y, z$ ($x \neq y \neq z) \in [1, n]$,

① when $\ell(x) + \ell(y) = \ell(z)$, there is $\ell(x) + \|W\| + \ell(y) + \|W\| \neq \ell(z) + \|W\|$ (% $\overline{M}$);

② when $\ell(x) + \ell(y) \neq \ell(z)$, there always exist
$$C_x \equiv A'_x W'^{\ell'(x)} (\% M), C_y \equiv A'_y W'^{\ell'(y)} (\% M), \text{ and } C_z \equiv A'_z W'^{\ell'(z)} (\% M)$$
such that $\ell'(x) + \ell'(y) \equiv \ell'(z)$ (% $\overline{M}$) with the constraint $A'_z \leq \mathcal{P}$.

Refer to [3] for its proof.

Notice that according to the proof of Property 5 in [3], it is not difficult to understand that when $\Omega = \{5, 6, \ldots, n+4\}$ is substituted with $\Omega = \{+/-5, +/-7, \ldots, +/-(2n+3)\}$, where "+/−" means the selection of the "+" or "−" sign, Property 5 still holds.

Property 5 illuminates that will continued fraction attack on $C_i \equiv A_i W^{\ell(i)}$ (% $M$) by Theorem 12.19 in Section 12.3 of [12] be utterly ineffectual only if elements in $\Omega$ are suitably selected [13].

## 3 Design of the New Cryptoscheme

Due to $L_p[1/3, 1.923] = 2^{80}$ with $p$ prime and $\lceil \lg p \rceil \approx 1024$ [14], the shortest bit-length of a plaintext block should be 80. In the new scheme, to acquire provable semantical security, random 16 bits are appended the terminal of a plaintext block of $n$ bits when it is encrypted.

Let $\tilde{n} = n + 16$ with $n = 80, 96,$ or $112$. Additionally, two adjacent bits are orderly treated as a unit, namely a bit-pair string $B_1\ldots B_{\tilde{n}/2}$ is utilized to represent a plaintext block $b_1\ldots b_{\tilde{n}} \neq 0$.

### 3.1 Key Generation Algorithm

Considering decryption speed, the absolute values of elements of $\Omega$ should be as small as possible, and every three successive elements of $\Omega$ are treated as a triple according to 2 bits to 3 items.

Let $\Lambda = \{2, \ldots, \mathcal{P}\}$, where $\mathcal{P} = 937, 991,$ or $1201$ corresponding to $\tilde{n} = 96, 112,$ or $128$ separately.

Let $\tilde{\iota} = \lceil \lg M \rceil = 464, 544,$ or $640$ corresponding to $\tilde{n} = 96, 112,$ or $128$ separately.

Assume that $\bar{A}_j$ is the maximum in $(A_{3j-2}, A_{3j-1}, A_{3j}) \; \forall j \in [1, \tilde{n}/2]$.

The following algorithm is generally employed by the owner of a key pair.

INPUT: the integer $n$; the integer $\tilde{\iota}$; the prime $\mathcal{P}$.

S1: Let $\tilde{n} \leftarrow n + 16$, $\Lambda \leftarrow \{2, \ldots, \mathcal{P}\}$;
yield the first $\tilde{n}$ primes $\dot{p}_1, \ldots, \dot{p}_{\tilde{n}}$ in the natural number set;
yield $\Omega \leftarrow \{(+/-(6j-1), +/-(6j+1), +/-(6j+3))_{\text{Æ}} \mid j = 1, \ldots, \tilde{n}/2\}$.

S2: Produce an odd coprime $\{A_1, \ldots, A_{3\tilde{n}/2} \mid A_i \in \Lambda\} = \{A_{3j-2}, A_{3j-1}, A_{3j} \mid j = 1, \ldots, \tilde{n}/2\}$;
arrange $\bar{A}_1, \ldots, \bar{A}_{\tilde{n}/2}$ to $\bar{A}_{x_1}, \ldots, \bar{A}_{x_{\tilde{n}/2}}$ in descending order.

S3: Find a prime $M > \bar{A}_{x_1}^{\tilde{n}/4+1} \prod_{i=2}^{\tilde{n}/4} \bar{A}_{x_i}$ making $\lceil \lg M \rceil = \tilde{\iota}$ and
$\prod_{i=1}^{k} \dot{p}_i^{e_i} \mid \bar{M}$, where $k$ meets $\prod_{i=1}^{k} e_i \geq 2^{10}$ and $\dot{p}_k < \tilde{n}$.

S4: Produce pairwise distinct $(\ell(3j-2), \ell(3j-1), \ell(3j)) \in \Omega$ for $j = 1, \ldots, \tilde{n}/2$.

S5: Stochastically pick $W, \delta \in (1, \bar{M})$ making $\|W\| \geq 2^{n-20}$ and $\gcd(\delta, \bar{M}) = 1$.

S6: Compute $C_i \leftarrow (A_i W^{\ell(i)})^{\delta} \% M$ for $i = 1, \ldots, 3\tilde{n}/2$.

OUTPUT: a public key $(\{C_1, \ldots, C_{3\tilde{n}/2}\}, M)$; a private key $(\{A_1, \ldots, A_{3\tilde{n}/2}\}, W, \delta, M)$.

The lever function $\{\ell(1), \ldots, \ell(\tilde{n}/2)\}$ is discarded but must not be divulged. Notice that

① at S1, $\Omega = \{(+/-(6j-1), +/-(6j+1), +/-(6j+3))_{\text{Æ}} \mid j = 1, \ldots, \tilde{n}/2\}$ indicates that $\Omega$ is one of $(3!)^{\tilde{n}/2} 2^{3\tilde{n}/2}$ potential sets consisting of 3-tuple elements, where "+/–" means the selection of the "+" or "–" sign, and the subscript Æ means that $(+/-(6j-1), +/-(6j+1), +/-(6j+3))_{\text{Æ}}$ is a permutation of $(+/-(6j-1), +/-(6j+1), +/-(6j+3))$;

② at S2, $\gcd(A_{3i-2}, A_{3i-1}, A_{3i}) \neq 1$ ($i \in [1, \tilde{n}/2]$) is allowed — $(3^3, 3^2, 3)$ for example since only one of three elements will occur in the product $\bar{G}$;

③ at S3, the inequation $M > \bar{A}_{x_1}^{\tilde{n}/4+1} \prod_{i=2}^{\tilde{n}/4} \bar{A}_{x_i}$ assures that a ciphertext can be decrypted correctly;

④ at S5, let $W \equiv g^{\bar{M}/F} (\% M)$, then $\|W\| = \bar{M} / \gcd(\bar{M}, \bar{M}/F)$ [11], where $F \geq 2^{n-20}$ is a factor of $\bar{M}$, and $g$ is a generator by Algorithm 4.80 in Section 4.6 of [14].

***Definition 6***: Given $(\{C_1, \ldots, C_{3\tilde{n}/2}\}, M)$, seeking the original $(\{A_1, \ldots, A_{3\tilde{n}/2}\}, \{\ell(1), \ldots, \ell(3\tilde{n}/2)\}, W, \delta)$ from $C_i \equiv (A_i W^{\ell(i)})^{\delta} (\% M)$ with $A_i \in \Lambda = \{2, \ldots, \mathcal{P} \mid \mathcal{P} \leq 1201\}$ and $\ell(i)$ from $\Omega = \{(+/-(6j-1), +/-(6j+1), +/-(6j+3))_{\text{P}} \mid j = 1, \ldots, \tilde{n}/2\}$ for $i = 1, \ldots, 3\tilde{n}/2$ is referred to as a multivariate permutation problem (MPP).

***Property 6***: The MPP $C_i \equiv (A_i W^{\ell(i)})^{\delta} (\% M)$ with $A_i \in \Lambda = \{2, \ldots, \mathcal{P} \mid \mathcal{P} \leq 1201\}$ and $\ell(i)$ from $\Omega = \{(+/-(6j-1), +/-(6j+1), +/-(6j+3))_{\text{P}} \mid j = 1, \ldots, \tilde{n}/2\}$ for $i = 1, \ldots, 3\tilde{n}/2$ is computationally at least equivalent to the DLP in the same prime field.

Refer to Section 4.1 of [3] for its proof.

## 3.2 Encryption Algorithm

This algorithm is employed by a person who wants to encrypt plaintexts.

INPUT: a public key $(\{C_1, \ldots, C_{3\tilde{n}/2}\}, M)$;
the bit-pair string $B_1 \ldots B_{n/2}$ of a plaintext block $b_1 \ldots b_n \neq 0$.

Notice that if the number of 00-pairs in $B_1 \ldots B_{n/2}$ is larger than $n/4$, let $b_1 \ldots b_n = \neg b_1 \ldots \neg b_n$ in order that a related ciphertext can be decrypted conforming to the constraint on $M$.

S1: Yield a random bit string $b_{n+1} \ldots b_{\tilde{n}}$ appended to $b_1 \ldots b_n$;
form $B_1 \ldots B_{\tilde{n}/2}$ until the number of 00-pairs $\leq \tilde{n}/4$.

S2: Set $C_0 \leftarrow 1$, $k \leftarrow 0$, $i \leftarrow 1$, $\bar{s} \leftarrow 0$.

S3: If $B_i = 00$ then let $k \leftarrow k + 1$, $B_i \leftarrow 0$
else {
let $B_i \leftarrow k + 1$, $k \leftarrow 0$;
if $\bar{s} = 0$ then $\bar{s} \leftarrow i$ else null.
}

S4: Let $i \leftarrow i + 1$;
if $i \leq \tilde{n}/2$ then goto S3.

S5: If $k \neq 0$ then let $B_{\bar{s}} \leftarrow B_{\bar{s}} + k$.

S6: Stochastically produce $r_1 \ldots r_{\tilde{n}/2} \in \{0, 1\}^{\tilde{n}/2}$;
set $r_{\bar{s}} \leftarrow 1$.

S7: Compute $\bar{G} \leftarrow \prod_{i=1}^{\tilde{n}/2} (C_{r_i(3(i-1)+B_i) + \neg r_i(3(i-B_i)+B_i)})^{B_i} \% M$.

OUTPUT: a ciphertext $\bar{G}$.

Evidently, a different ciphertext will be outputted every time an identical plaintext is inputted repeatedly. The identical plaintext may correspond to at most $2^{\tilde{n}/4}2^{\tilde{n}-n}$ different ciphertexts since $\tilde{n}/2$ bit-pairs may be interlaced with a 00-pair and a non-00-pair, and $b_{n+1}\ldots b_{\tilde{n}}$ is produced randomly. It will take the running time of $O(\tilde{n}2^{\tilde{n}/2}2^{\tilde{n}-n}\lg^2 M)$ bit operations exhaustively to search all the possible ciphertexts of a plaintext.

Note a JUOAN ciphertext $\prod_{i=1}^{\tilde{n}/2}(C_{r_i(3(i-1)+B_i) + \neg r_i(3(i-B_i)+B_i)})^{B_i}(\% M)$ is different from a Naccache-Stern ciphertext $c \equiv \prod_{i=1}^{n} v_i^{b_i}(\% M)$ [15], where $v_i \equiv \dot{\rho}_i^{1/\delta}$ (% $M$) with $\dot{\rho}_i$ prime is a public key.

***Definition 7:*** Given $(\{C_1, \ldots, C_{3\tilde{n}/2}\}, M)$ and $\bar{G}$, seeking $B_1\ldots B_{\tilde{n}/2}$ from $\bar{G}_1 \equiv \prod_{i=1}^{\tilde{n}/2}(C_{3(i-1)+B_i})^{\lceil B/3 \rceil}$ (% $M$) with $C_0 = 1$ is referred to as a subset product problem (SPP), where $B_1\ldots B_{\tilde{n}/2}$ is the bit-pair string of $b_1\ldots b_{\tilde{n}} \neq 0$.

***Property 7:*** The SPP $\bar{G}_1 \equiv \prod_{i=1}^{\tilde{n}/2}(C_{3(i-1)+B_i})^{\lceil B/3 \rceil}$ (% $M$) with $C_0 = 1$ is computationally at least equivalent to the DLP in the same prime field, where $B_1\ldots B_{\tilde{n}/2} \neq 0$ is a bit-pair string.

***Definition 8:*** Given $(\{C_1, \ldots, C_{3\tilde{n}/2}\}, M)$ and $\bar{G}$, seeking $B_1\ldots B_{\tilde{n}/2}$ from $\bar{G} \equiv \prod_{i=1}^{\tilde{n}/2}(C_{3(i-1)+B_i})^{B_i}$ (% $M$) or $\bar{G} \equiv \prod_{i=1}^{\tilde{n}/2}(C_{r_i(3(i-1)+B_i) + \neg r_i(3(i-B_i)+B_i)})^{B_i}$ (% $M$) with $C_0 = 1$ and $r_1\ldots r_{\tilde{n}/2}$ a random bit string is referred to as an anomalous subset product problem (ASPP), where $\mathcal{B}_1\ldots \mathcal{B}_{\tilde{n}/2}$ is the bit-pair shadow string of $B_1\ldots B_{\tilde{n}/2}$ corresponding to $b_1\ldots b_{\tilde{n}} \neq 0$.

***Property 8:*** The ASPP $\bar{G} \equiv \prod_{i=1}^{\tilde{n}/2}(C_{3(i-1)+B_i})^{\mathcal{B}_i}$ (% $M$) with $C_0 = 1$ is computationally at least equivalent to the DLP in the same prime field, where $\mathcal{B}_1\ldots \mathcal{B}_{\tilde{n}/2}$ is the bit-pair shadow string of $B_1\ldots B_{\tilde{n}/2} \neq 0$.

***Property 9:*** The ASPP $\bar{G} \equiv \prod_{i=1}^{\tilde{n}/2}(C_{r_i(3(i-1)+B_i) + \neg r_i(3(i-B_i)+B_i)})^{\mathcal{B}_i}$ (% $M$) with $C_0 = 1$ and $r_1\ldots r_{\tilde{n}/2}$ a random bit string is computationally at least equivalent to the DLP in the same prime field, where $\mathcal{B}_1\ldots \mathcal{B}_{\tilde{n}/2}$ is the bit-pair shadow string of $B_1\ldots B_{\tilde{n}/2} \neq 0$.

See Section 6 for the proofs of Property 7, 8, and 9.

## 3.3 Decryption Algorithm

This algorithm is employed by a person who holds a private key and wants to decrypt ciphertexts.

INPUT: a private key $(\{A_1, \ldots, A_{3\tilde{n}/2}\}, W, \delta, M)$; a ciphertext $\bar{G}$.

It should be noted that due to $2 | \sum_{i=1}^{\tilde{n}/2} \mathcal{B}_i$ and $2 | \ell(r_i(3(i-1)+B_i)+\neg r_i(3(i-B_i)+B_i))$ for $i \in [1, \tilde{n}/2]$ with $\ell(0) = 0$, $\underline{k} = \sum_{i=1}^{\tilde{n}/2} \mathcal{B}_i \ell(r_i(3(i-1)+B_i)+\neg r_i(3(i-B_i)+B_i))$ must be even.

S1: Compute $Z_0 \leftarrow \bar{G}^{\delta^{-1}} \% M$; set $Z_1 \leftarrow Z_0$, $\varsigma \leftarrow 0$.

S2: If $2 | Z_\varsigma$ then {do $Z_\varsigma \leftarrow Z_\varsigma W^{2^{(-1)^\varsigma}} \% M$; goto S2.}

S3: Set $B_1\ldots B_{n/2} \leftarrow 0$, $j \leftarrow 0$, $k \leftarrow 0$, $\tilde{v} \leftarrow 0$, $i \leftarrow 1$, $G \leftarrow Z_\varsigma$.

S4: If $(A_{3i-j})^{\tilde{v}+1} | G$ then {let $\tilde{v} \leftarrow \tilde{v} + 1$; goto S4.}

S5: Let $j \leftarrow j + 1$; if $\tilde{v} = 0$ and $j \leq 2$ then goto S4.

S6: If $\tilde{v} = 0$ then

    S6.1: let $k \leftarrow k + 1$, $i \leftarrow i + 1$

  else

    S6.2: compute $G \leftarrow G / (A_{3i-j})^{\tilde{v}}$;

    S6.3: if $k > 0$ or $\tilde{v} \geq i$ then let $B_i \leftarrow 3 - j$, $i \leftarrow i + 1$

        else let $B_{i+\tilde{v}-1} \leftarrow 3 - j$, $i \leftarrow i + \tilde{v}$;

    S6.4: set $\tilde{v} \leftarrow 0$, $k \leftarrow 0$.

S7: If $i \leq \tilde{n}/2$ and $G \neq 1$ then set $j \leftarrow 0$, goto S4.

S8: If $G \neq 1$ then {set $\varsigma \leftarrow \neg \varsigma$; do $Z_\varsigma \leftarrow Z_\varsigma W^{2^{(-1)^\varsigma}} \% M$; goto S2.}

S9: Extract $B_1\ldots B_{n/2}$ from $B_1\ldots B_{\tilde{n}/2}$.

OUTPUT: a related plaintext $B_1\ldots B_{n/2}$, namely $b_1\ldots b_n$.

Notice that only if $\bar{G}$ is a true ciphertext, can the algorithm always terminate normally, and $b_1\ldots b_n$ will be original although $r_1\ldots r_{\tilde{n}/2}$ is brought into an encryption process.

### 3.4 Running Times of the Algorithms

The running time of an algorithm is measured in the number of bit operations [14], and it has an asymptotic implication. According to [14], a modular addition will take $O(\lg M)$ bit operations (shortly, bos), and a modular multiplication will take $O(2\lg^2 M)$ bos.

#### 3.4.1 Running Time of the Key Generator

In the key generator, the steps which exert dominant effects on running time are S5 and S6.

At S5, seeking $W$ namely seeking $g$ will take $O(6(\ln\ln \bar{M})\lg^3 M)$ bos [14]. For every $i$, S6 contains one modular power $(A_i W^{\ell(i)})^\delta$, where $W^{\ell(i)}$ will take $2\lg(3\tilde{n}/2)$ modular multiplications which is subject to one modular power since $\tilde{n}$ is far smaller than $M$. Thus, the running time of the key generator is about $O((6\ln\ln \bar{M} + 4(3\tilde{n}/2))\lg^3 M) = O(6(\ln\ln \bar{M} + \tilde{n})\lg^3 M)$ bos.

#### 3.4.2 Running Time of the Encryption Algorithm

In the encryption algorithm, the dominant step is S7.

Due to $\sum_{i=1}^{\tilde{n}/2} B_i = \tilde{n}/2$, S7 contains at most $\tilde{n}/2$ modular multiplications. Thus, the running time of the encryption algorithm is $O(\tilde{n}\lg^2 M)$ bos.

#### 3.4.3 Running Time of the Decryption Algorithm

In the decryption algorithm, the dominant step is S2 which pairs with S8 to form a loop. S1 of a modular power is subject to the loop S2 ↔ S8.

It is easy to see that the number of times of executing the loop S2 ↔ S8 which mainly contains a modular multiplication is $\underline{k} = \sum_{i=1}^{\tilde{n}/2} B_i \ell(r_i(3(i-1)+B_i) + \neg r_i(3(i-B_i)+B_i))$, where $B_i$ is relevant to a plaintext block, and $\ell(r_i(3(i-1)+B_i)+\neg r_i(3(i-B_i)+B_i))$ is relevant to the set $\Omega = \{(+/-(6j-1), +/-(6j+1), +/-(6j+3))_P \mid j=1, \ldots, \tilde{n}/2\}$ which is indeterminate. Therefore, it is very difficult accurately to know the value of $\underline{k}$.

When a plaintext block $B_1 \ldots B_{\tilde{n}/2}$ contains $\tilde{n}/4$ successive 00-pairs, we may obtain the maximal value of $\underline{k}$ on condition that every $\ell(r_i(3(i-1)+B_i)+\neg r_i(3(i-B_i)+B_i))$ has the same operating sign.

The maximal value of $\underline{k}$ is

$$\underline{k} = (\tilde{n}/4+1)(2(3\tilde{n}/2)+3) + (2(3\tilde{n}/2)-3) + (2(3\tilde{n}/2)-9) + \ldots$$
$$+(2(3\tilde{n}/2)+3 - 6(\tilde{n}/4-1))$$
$$= (3/4)(\tilde{n}+4)(\tilde{n}+1) + (3/16)(3\tilde{n}+4)(\tilde{n}-4)$$
$$= (3/16)\tilde{n}(7\tilde{n}+12).$$

Considering that $\ell(r_i(3(i-1)+B_i)+\neg r_i(3(i-B_i)+B_i))$ may be positive or negative, the minimal value of $\underline{k}$ will be 0 with a suitable plaintext block $B_1 \ldots B_{\tilde{n}/2}$.

Thus, the simply expected value of $\underline{k}$ is $(3/32)\tilde{n}(7\tilde{n}+12) \approx (21/32)\tilde{n}^2$.

Again considering that $W^{-2}$ and $W^2$ is multiplied alternately every time, the simply expected value of $\underline{k}$ should be $2(21/32)\tilde{n}^2 \approx \tilde{n}^2$.

In summary, the simply expected time of the decryption algorithm is $O(2\tilde{n}^2 \lg^2 M)$ bos.

Practically, because the possible values of $\underline{k}$ (including the repeated) will distribute in the integral interval $[0, (3/16)\tilde{n}(7\tilde{n}+12)]$ which includes the points $0, 2, 4, \ldots, (3/16)\tilde{n}(7\tilde{n}+12)$, and is the maximal possible range, and the probability that $\underline{k}$ takes large integers is comparatively small, the concrete running time of a decryption will be far smaller than $O(2\tilde{n}^2 \lg^2 M)$ bos.

## 4 Correctness and Uniqueness

### 4.1 Correctness of the Decryption Algorithm

Since $(\mathbb{Z}_M^*, \cdot)$ is an Abelian group, namely a commutative group, $\forall \underline{k} \in [1, \bar{M}]$ there is $W^{\underline{k}}(W^{-1})^{\underline{k}} \equiv W^{\underline{k}} (W^{\underline{k}})^{-1} \equiv 1 \;(\% M)$, where $W \in [1, \bar{M}]$ is any arbitrary integer.

**Fact 3**: Let $\underline{k} = \sum_{i=1}^{\tilde{n}/2} B_i \ell(r_i(3(i-1)+B_i)+\neg r_i(3(i-B_i)+B_i)) \;\% \bar{M}$ with $\ell(0) = 0$, where $B_1 \ldots B_{\tilde{n}/2}$ is the bit-pair shadow string of $B_1 \ldots B_{\tilde{n}/2}$ corresponding to $b_1 \ldots b_{\tilde{n}} \neq 0$, and $r_1 \ldots r_{\tilde{n}/2}$ is a random bit string. Then $\hat{G}^{\delta^{-1}}(W^{-1})^{\underline{k}} \equiv \prod_{i=1}^{\tilde{n}/2} (A_{r_i(3(i-1)+B_i) + \neg r_i(3(i-B_i)+B_i)})^{B_i} \;(\% M)$.

*Proof*:

Let $b_1 \ldots b_{\tilde{n}}$, namely $B_1 \ldots B_{\tilde{n}/2}$ be a plaintext of $\tilde{n}$ bits. Additionally, let $A_0 = 1$.

According to the key generator, the encryption algorithm, and $\sum_{i=1}^{\tilde{n}/2} B_i = \tilde{n}/2$, there is

$$\bar{G} \equiv \prod_{i=1}^{\tilde{n}/2} (C_{r_i(3(i-1)+B_i) + \neg r_i(3(i-B_i)+B_i)})^{B_i}$$

$$\equiv \prod_{i=1}^{\tilde{n}/2} ((A_{r_i(3(i-1)+B_i) + \neg r_i(3(i-B_i)+B_i)} W^{\ell(r_i(3(i-1)+B_i) + \neg r_i(3(i-B_i)+B_i))})^{\delta})^{B_i}$$

$$\equiv W^{(\sum_{i=1}^{\tilde{n}/2} B_i \ell(r_i(3(i-1)+B_i) + \neg r_i(3(i-B_i)+B_i)))\delta} \prod_{i=1}^{\tilde{n}/2} (A_{r_i(3(i-1)+B_i) + \neg r_i(3(i-B_i)+B_i)})^{\delta B_i}$$

$$\equiv W^{\underline{k}\delta} \prod_{i=1}^{\tilde{n}/2} (A_{r_i(3(i-1)+B_i) + \neg r_i(3(i-B_i)+B_i)})^{\delta B_i} \ (\% \ M).$$

Further, raising either side of the above congruence to the $\delta^{-1}$-th yields

$$\bar{G}^{\delta^{-1}} \equiv (W^{\underline{k}\delta} \prod_{i=1}^{\tilde{n}/2} (A_{r_i(3(i-1)+B_i) + \neg r_i(3(i-B_i)+B_i)})^{\delta B_i})^{\delta^{-1}}$$

$$\equiv W^{\underline{k}} \prod_{i=1}^{\tilde{n}/2} (A_{r_i(3(i-1)+B_i) + \neg r_i(3(i-B_i)+B_i)})^{B_i} \ (\% \ M).$$

Multiplying either side of the just above congruence by $(W^{-1})^{\underline{k}}$ yields

$$\bar{G}^{\delta^{-1}} (W^{-1})^{\underline{k}} \equiv W^{\underline{k}} \prod_{i=1}^{\tilde{n}/2} (A_{r_i(3(i-1)+B_i) + \neg r_i(3(i-B_i)+B_i)})^{B_i} (W^{-1})^{\underline{k}}$$

$$\equiv \prod_{i=1}^{\tilde{n}/2} (A_{r_i(3(i-1)+B_i) + \neg r_i(3(i-B_i)+B_i)})^{B_i} \equiv G \ (\% \ M).$$

Clearly, the above process also gives a method of seeking $G$ at one time. □

Notice that in practice, $\underline{k}$ is unknowable in advance.

However, since $|\underline{k}| < \tilde{n}(2(3\tilde{n}/2) + 3)/2 = 3\tilde{n}(\tilde{n}+1)/2$ is comparatively small, we may search $\underline{k}$ heuristically by multiplying $W^{-2}$ or $W^2 \ \% \ M$ and judging whether $G = 1$ after it is divided exactly by some $(A_{3i-j})^{\tilde{v}}$. It is known from the decryption algorithm that the original $B_1 \ldots B_{\tilde{n}/2}$ will be acquired at the same time the condition $G = 1$ is satisfied.

### 4.2 Uniqueness of a Plaintext Solution

Because the public key $\{C_1, \ldots, C_{3\tilde{n}/2}\}$ is a non-coprime sequence, the mapping from $B_1 \ldots B_{\tilde{n}/2}$ to $\bar{G} = \prod_{i=1}^{\tilde{n}/2} (C_{r_i(3(i-1)+B_i) + \neg r_i(3(i-B_i)+B_i)})^{B_i} \ \% \ M$ is theoretically many-to-one. It might possibly result in the nonuniqueness of a plaintext solution $B_1 \ldots B_{\tilde{n}/2}$ when $\bar{G}$ is being unveiled.

*Fact 4*: The probability that a solution for $\bar{G} = \prod_{i=1}^{\tilde{n}/2} (C_{r_i(3(i-1)+B_i) + \neg r_i(3(i-B_i)+B_i)})^{B_i} \ \% \ M$ is nonunique is nearly zero, where the solution is namely a plaintext.

*Proof*:

Suppose that a ciphertext $\bar{G}$ can be obtained respectively from two different bit-pair strings $B_1 \ldots B_{\tilde{n}/2}$ and $B'_1 \ldots B'_{\tilde{n}/2}$. Then,

$$\bar{G} \equiv \prod_{i=1}^{\tilde{n}/2} (C_{r_i(3(i-1)+B_i) + \neg r_i(3(i-B_i)+B_i)})^{B_i} \equiv \prod_{i=1}^{\tilde{n}/2} (C_{r_i(3(i-1)+B'_i) + \neg r_i(3(i-B'_i)+B'_i)})^{B'_i} \ (\% \ M).$$

That is,

$$\prod_{i=1}^{\tilde{n}/2} (A_{r_i(3(i-1)+B_i) + \neg r_i(3(i-B_i)+B_i)} W^{\ell(r_i(3(i-1)+B_i) + \neg r_i(3(i-B_i)+B_i))})^{\delta B_i}$$

$$\equiv \prod_{i=1}^{\tilde{n}/2} (A_{r_i(3(i-1)+B'_i) + \neg r_i(3(i-B'_i)+B'_i)} W^{\ell(r_i(3(i-1)+B'_i) + \neg r_i(3(i-B'_i)+B'_i))})^{\delta B'_i} \ (\% \ M).$$

Further, there is

$$W^{\underline{k}\delta} \prod_{i=1}^{\tilde{n}/2} (A_{r_i(3(i-1)+B_i) + \neg r_i(3(i-B_i)+B_i)})^{\delta B_i} \equiv W^{\underline{k}'\delta} \prod_{i=1}^{\tilde{n}/2} (A_{r_i(3(i-1)+B'_i) + \neg r_i(3(i-B'_i)+B'_i)})^{\delta B'_i} (\% M),$$

where $\underline{k} = \sum_{i=1}^{\tilde{n}/2} B_i \ell(r_i(3(i-1)+B_i) + \neg r_i(3(i-B_i)+B_i))$, and $\underline{k}' = \sum_{i=1}^{\tilde{n}/2} B'_i \ell(r_i(3(i-1)+B'_i) + \neg r_i(3(i-B'_i)+B'_i)) \ \% \ \bar{M}$ with $\ell(0) = 0$.

Raising either side of the above congruence to the $\delta^{-1}$-th power yields

$$W^{\underline{k}} \prod_{i=1}^{\tilde{n}/2} (A_{r_i(3(i-1)+B_i) + \neg r_i(3(i-B_i)+B_i)})^{B_i} \equiv W^{\underline{k}'} \prod_{i=1}^{\tilde{n}/2} (A_{r_i(3(i-1)+B'_i) + \neg r_i(3(i-B'_i)+B'_i)})^{B'_i} (\% M).$$

Without loss of generality, let $\underline{k} \geq \underline{k}'$. Because $(\mathbb{Z}_M^*, \cdot)$ is an Abelian group, there is

$$W^{\underline{k}-\underline{k}'} \equiv \prod_{i=1}^{\tilde{n}/2} (A_{r_i(3(i-1)+B'_i) + \neg r_i(3(i-B'_i)+B'_i)})^{B'_i} (A_{r_i(3(i-1)+B_i) + \neg r_i(3(i-B_i)+B_i)})^{-B_i} \ (\% \ M).$$

Let $\theta \equiv \prod_{i=1}^{\tilde{n}/2} (A_{r_i(3(i-1)+B'_i) + \neg r_i(3(i-B'_i)+B'_i)})^{B'_i} (A_{r_i(3(i-1)+B_i) + \neg r_i(3(i-B_i)+B_i)})^{-B_i} \ (\% \ M)$, namely

$$\theta \equiv W^{\underline{k}-\underline{k}'} \ (\% \ M).$$

This congruence signifies when the plaintext $B_1 \ldots B_{\tilde{n}/2}$ is not unique, the value of $W$ must be relevant to $\theta$. The contrapositive assertion equivalent to it is that if the value of $W$ is irrelevant to $\theta$, $B_1 \ldots B_{\tilde{n}/2}$ will be unique. Thus, we need to consider the probability that $W$ takes a value relevant to $\theta$.

If an adversary tries to attack an 80-bit symmetric key through the exhaustive search, and a computer can verify trillion values per second, then it will take 38334 years for the adversary to verify all the potential values. Hence, currently 80 bits are quite enough for the security of a symmetric key.

$B_1...B_{\tilde{n}/2}$ contains $\tilde{n}$ bits which indicates $\prod_{i=1}^{\tilde{n}/2}(C_{r_i(3(i-1)+B_i)+\rightarrow r_i(3(i-\bar{B}_i)+B_i)})^{\bar{B}_i}$ has $2^{\tilde{n}}$ potential values, and thus the number of potential values of $\theta$ is at most $2^{\tilde{n}} \times 2^{\tilde{n}}$. Notice that since $A_1^{-1}, ..., A_{3\tilde{n}/2}^{-1}$ are not necessarily coprime, some values of $\theta$ may possibly occur repeatedly.

Because $|k - k'| < 3\tilde{n}(\tilde{n}+1) \le 47601 \approx 2^{16}$ with $\tilde{n} \le 128$, and $W$ has at most $2^{16}$ solutions to every value of $\theta$, the probability that $W$ takes a value relevant to $\theta$ is at most $2^{16}2^{2\tilde{n}}/M$.

When $\tilde{n} \ge 96$, there is $2^{16}2^{2\tilde{n}}/M \le 2^{208}/2^{464} = 1/2^{256}$, and close to zero, where $M$ is subject to $\lceil \lg M \rceil = 464, 544,$ or $640$ corresponding with $\tilde{n} = 96, 112,$ or $128$. □

## 5 Security Analysis of a Private Key

In the below cryptanalysis, we suppose that the integer factorization problem (IFP) $N=pq$ with $\lceil \lg N \rceil < 1024$ [4], the discrete logarithm problem (DLP) $y \equiv g^x$ (% $p$) with $\lceil \lg p \rceil < 1024$ [16][17], and the subset sum problem (SSP) of low density $s \equiv \sum_{i=1}^{n} c_i b_i$ (% $M$) with $D \approx n / \lceil \lg M \rceil < 1$ and $n < \lceil \lg M \rceil < 1024$ [9] can be solved in tolerable subexponential time or in polynomial time [18][19].

Notice that the structure of the set $\Omega$ consisting of triples has no change in essence compared with the $\Omega$ in [3].

### 5.1 Attack by Interaction of the Key Transform Items

In the key transform $C_i \equiv (A_i W^{\ell(i)})^\delta$ (% $M$), the parameters $A_i \in \Lambda = \{2, 3, ..., Þ | Þ = 937, 991,$ or $1201\}$ and $\ell(i)$ from $\Omega = \{(+/-(6j-1), +/-(6j+1), +/-(6j+3))_Æ | j = 1, ..., \tilde{n}/2\}$ seem vulnerable.

### 5.1.1 Eliminating $W$ through $\ell(x_1) + \ell(x_2) = \ell(y_1) + \ell(y_2)$

$\forall x_1, x_2, y_1, y_2 \in [1, 3\tilde{n}/2]$, assume that $\ell(x_1) + \ell(x_2) = \ell(y_1) + \ell(y_2)$.
Let $G_z \equiv C_{x_1} C_{x_2} (C_{y_1} C_{y_2})^{-1}$ (% $M$), namely
$$G_z \equiv (A_{x_1} A_{x_2} (A_{y_1} A_{y_2})^{-1})^\delta \text{ (% } M\text{)}.$$

If an adversary divines the values of $A_{x_1}, A_{x_2}, A_{y_1}, A_{y_2} \in \Lambda$, he may compute $\delta$ through a discrete logarithm in $L_M[1/3, 1.923]$ time, where $M < 2^{640}$.

However, a concrete $\Omega$ is one of $(2^3 3!)^{\tilde{n}/2}$ potential sets, indeterminate, and unknown due to $|\Omega| = \tilde{n}/2$ and $\Omega = \{(+/-(6j-1), +/-(6j+1), +/-(6j+3))_P | j=1, ..., \tilde{n}/2\}$.

For example, assume that $\ell(x_1) + \ell(x_2) = 5 + 11$, and $\ell(y_1) + \ell(y_2) = -7 + 9$, then there is $\ell(x_1) + \ell(x_2) \ne \ell(y_1) + \ell(y_2)$. Therefore, among $\ell(1), ...,$ and $\ell(3\tilde{n}/2)$, there does not necessarily exist $\ell(x_1) + \ell(x_2) = \ell(y_1) + \ell(y_2)$.

The above example illustrates that to determinate the existence of $\ell(x_1) + \ell(x_2) = \ell(y_1) + \ell(y_2)$, the adversary must first determinate the constitution of the set $\Omega$, which will have $O((2^3 3!)^{\tilde{n}/2})$ time complexity.

### 5.1.2 Eliminating $W$ through the $\|W\|$-th Power

Owing to $\lceil \lg M \rceil = 544$ or $640$, $\bar{M}$ can be factorized in tolerable subexponential time. Again owing to $\prod_{i=1}^{k} \dot{p}_i^{e_i} | \bar{M}$ and $\prod_{i=1}^{k} e_i \ge 2^{10}$ with $\dot{p}_k < \tilde{n}$, $\|W\|$ can be divined in the time of about the $2^{10}$ magnitude.

Raising either side of $C_i \equiv (A_i W^{\ell(i)})^\delta$ % $M$ to the $\|W\|$-th power yields
$$C_i^{\|W\|} \equiv (A_i)^{\delta \|W\|} \text{ (% } M\text{)}.$$

Let $C_i \equiv g^{u_i}$ (% $M$), and $A_i \equiv g^{v_i}$ (% $M$), where $g$ is a generator of $(\mathbb{Z}_M^*, \cdot)$. Then
$$u_i \|W\| \equiv v_i \|W\| \delta \text{ (% } \bar{M}\text{)}$$
for $i = 1, ..., 3\tilde{n}/2$. Notice that $u_i \ne v_i \delta$ (% $\bar{M}$) due to $\|W\| | \bar{M}$.

The above congruence looks to be the MH transform [9]. Actually, $\{v_1\|W\|, ..., v_{3\tilde{n}/2}\|W\|\}$ is not a super increasing sequence, and moreover there is not necessarily $\lg(u_i\|W\|) = \lg \bar{M}$.

Because $v_i\|W\| \in [1, \bar{M}]$ is stochastic, the inverse $\delta^{-1}$ % $\bar{M}$ not need be close to the minimum

$\bar{M}/(u_i\|W\|)$, $2\bar{M}/(u_i\|W\|)$, …, or $(u_i\|W\| - 1)\bar{M}/(u_i\|W\|)$. Namely $\delta^{-1}$ may lie at any integral position in the interval $[k\bar{M}/(u_i\|W\|), (k+1)\bar{M}/(u_i\|W\|)]$, where $k = 0, 1, …, u_i\|W\| - 1$, which illustrates the accumulation points of minima do not exist. Further observing, in this case, when $i$ traverses the interval $[2, 3\tilde{n}/2]$, the number of intersections of the intervals including $\delta^{-1}$ is likely the max of $(u_2\|W\|, …, u_{3\tilde{n}/2}\|W\|)$ which is promisingly close to $\bar{M}$. Therefore, the Shamir attack by the accumulation point of minima is fully ineffectual [20].

Even though find out $\delta^{-1}$ by the Shamir attack method, because each of $v_i$ has $\|W\|$ solutions, the number of potential sequences $\{g^{v_1}, …, g^{v_{3\tilde{n}/2}}\}$ is up to $\|W\|^{3\tilde{n}/2}$. Due to needing to verify whether $\{g^{v_1}, …, g^{v_{3\tilde{n}/2}}\}$ is a coprime sequence for each different sequence $\{v_1, …, v_{3\tilde{n}/2}\}$, the number of coprime sequences is in direct proportion to $\|W\|^{3\tilde{n}/2}$. Hence, the initial $\{A_1, …, A_{3\tilde{n}/2}\}$ cannot be determined in subexponential time. Further, the value of $W$ cannot be computed, and the values of $\|W\|$ and $\delta^{-1}$ cannot be verified in subexponential time, which indicates that MPP can also be resistant to the attack by the accumulation point of minima.

Additionally, an adversary may divine value of $A_i$ in about $|\varLambda|$ time, where $i \in [1, 3\tilde{n}/2]$, and compute $\delta$ by $u_i\|W\| \equiv v_i\|W\|\delta \ (\% \ \bar{M})$.

However, Owing to $\|W\| \mid \bar{M}$, the equation will have $\|W\|$ solutions. Therefore, the time complexity of finding the original $\delta$ is at least

$$\mathcal{T}_e = (3\tilde{n}/2)|\varLambda|L_M[1/3, 1.923] + 2^{10}|\varLambda|\|W\|$$
$$= 2^9(3\tilde{n})L_M[1/3, 1.923] + 2^{10}2^{10}2^{n-20}$$
$$\approx 2^9(3\tilde{n})L_M[1/3, 1.923] + 2^n > 2^n.$$

It is exponential in $n$ with $n = 80, 96$, or $112$.

## 5.2 Attack by a Certain Single $C_i$

Assume that there is only a solitary $C_i = (A_i W^{\ell(i)})^\delta \% M$ — $i = 1$ for example, and other $C_i$'s ($i = 2, …, 3\tilde{n}/2$) are unknown for adversaries.

Through divining the values of $A_1 \in \varLambda$, $\ell(1)$ from $\varOmega$, and $\delta$ coprime to $\bar{M}$, the secrete parameters $W \in (1, \bar{M})$ can be computed. Thus, the number of possible values of $W$ will be larger than $|\varOmega||\varLambda|(\bar{M}/\ln\bar{M}) > 2^{\tilde{n}}$, which manifests that the original $(A_1, \ell(1), W, \delta)$ cannot be determined in subexponential time.

Evidently, if $g_1 \equiv A_1 W^{\ell(1)} \ (\% \ M)$ is a constant, solving $C_1 = g_1^\delta \% M$ for $\delta$ is equivalent to the DLP. Factually, $g_1$ is not a constant, and at present, the time complexity of seeking the original $g_1$, namely $A_1 W^{\ell(1)}$ will be $O(M) > O(2^{\tilde{n}})$.

In summary, the time complexity of inferring a related private key from a public key is at least $O(2^{\tilde{n}})$.

# 6 Security Analysis of a Plaintext

The security of a plaintext depends on the ASPP $\bar{G} \equiv \prod_{i=1}^{\tilde{n}/2}(C_{r_i(3(i-1)+B_i)+\neg r_i(3(i-B_i)+B_i)})^{B_i} \ (\% \ M)$ with $C_0 = 1$ and $r_1…r_{\tilde{n}/2}$ a random bit string.

***Definition 9***: Let $A$ and $B$ be two computational problems. $A$ is said to reduce to $B$ in polynomial time, written as $A \leq_T^P B$, if there is an algorithm for solving $A$ which calls, as a subroutine, a hypothetical algorithm for solving $B$, and runs in polynomial time, excluding the time of the algorithm for $B$ [14][21].

***Definition 10***: Let $A$ and $B$ be two computational problems. If $A \leq_T^P B$ and $B \leq_T^P A$, then $A$ and $B$ are said to be computationally equivalent, written as $A =_T^P B$ [14][21].

Definition 9 and 10 suggest a reductive proof method called polynomial time Turing reduction (PTR) [21].

Naturally, we will enquire whether $A <_T^P B$ exists or not. The definition of $A <_T^P B$ may possibly be given theoretically, but the proof of $A <_T^P B$ is not easy in practice.

Let $\hat{H}(y = f(x))$ represent the complexity or hardness of solving the problem $y = f(x)$ for $x$ [18].

## 6.1 Proofs of Three Properties

- The proof of Property 7.

*Proof:*

For clear explanations, we extend $B_1…B_{\tilde{n}/2}$ to a bit string $b'_1…b'_{3\tilde{n}/2}$ by the following rule for $i = 1, …, \tilde{n}/2$:

① when $B_i = 0$, let $b'_{3(i-1)+1} = b'_{3(i-1)+2} = b'_{3(i-1)+3} = 0$;

② when $B_i \neq 0$, let $b'_{3(i-1)+1} = b'_{3(i-1)+2} = b'_{3(i-1)+3} = 0$, and $b'_{3(i-1)+B_i} = 1$.

Hence, we have the equivalent $\bar{G}_1 \equiv \prod_{i=1}^{3\tilde{n}/2} C_i^{b'_i}$ (% $M$).

The form of $\bar{G}_1$ here is similar to that of $\bar{G}_1$ in [3].

Especially, define $\bar{G}_1 \equiv \prod_{i=1}^{3\tilde{n}/2} C^{2^{3\tilde{n}/2-i} b'_i} \equiv \prod_{i=1}^{3\tilde{n}/2} (C^{2^{3\tilde{n}/2-i}})^{b'_i}$ (%$M$) when $C_1 = … = C_{3\tilde{n}/2} = C$.

Obviously, $\prod_{i=1}^{3\tilde{n}/2} C_i^{b'_i} = LM + \bar{G}_1$. Owing to $L \in [1, \overline{M}]$, deriving the non-modular product $\prod_{i=1}^{3\tilde{n}/2} C_i^{b'_i}$ from $\bar{G}_1$ is infeasible, which means inferring $b'_1…b'_{3\tilde{n}/2}$ from $\bar{G}_1$ is not a factorization problem.

Assume that $\bar{O}_s(\bar{G}_1, C_1, …, C_{3\tilde{n}/2}, M)$ is an oracle on solving $\bar{G}_1 \equiv \prod_{i=1}^{3\tilde{n}/2} C_i^{b'_i}$ (%$M$) for $b'_1…b'_{3\tilde{n}/2}$.

Let $y \equiv g^x$ (% $M$) be of the DLP, where $g$ is a generator of $(\mathbb{Z}_M^*, \cdot)$, and the binary form of $x$ is $b_1…b_{3\tilde{n}/2}$, namely

$$y \equiv \prod_{i=1}^{3\tilde{n}/2} (g^{2^{3\tilde{n}/2-i}})^{b_i} (\% \ M).$$

Then, by calling $\bar{O}_s(y, g^{2^{3\tilde{n}/2-1}}, …, g, M)$, $b_1…b_{3\tilde{n}/2}$ namely $x$ can be found.

According to Definition 9, there is

$$\hat{H}(y \equiv g^x (\% \ M)) \leq_T^P \hat{H}(\bar{G}_1 \equiv \prod_{i=1}^{\tilde{n}/2} (C_{3(i-1)+B_i})^{\lceil B_i/3 \rceil} (\% \ M)),$$

namely the SPP is at least equivalent to the DLP in the same prime field in complexity. □

- The proof of Property 8.

*Proof:*

For clear explanations, we extend $B_1…B_{\tilde{n}/2}$ to a nonrigid shadow string $b'_1…b'_{3\tilde{n}/2}$ by the following rule for $i = 1, …, \tilde{n}/2$:

① when $B_i = 0$, let $b'_{3(i-1)+1} = b'_{3(i-1)+2} = b'_{3(i-1)+3} = 0$;

② when $B_i \neq 0$, let $b'_{3(i-1)+1} = b'_{3(i-1)+2} = b'_{3(i-1)+3} = 0$, and $b'_{3(i-1)+B_i} = B_i$.

Hence, we have the equivalent

$$\bar{G} \equiv \prod_{i=1}^{3\tilde{n}/2} C_i^{b'_i} (\% \ M).$$

The form of $\bar{G}$ here is similar to that of $\bar{G}$ in [3].

Assume that $\bar{O}_a(\bar{G}, C_1, …, C_{3\tilde{n}/2}, M)$ is an oracle on solving $\bar{G} \equiv \prod_{i=1}^{3\tilde{n}/2} C_i^{b'_i}$ (% $M$) for $b'_1…b'_{3\tilde{n}/2}$, where $b'_1…b'_{3\tilde{n}/2}$ is a nonrigid shadow string corresponding to $B_1…B_{\tilde{n}/2}$.

Especially, define

$$\bar{G} \equiv \prod_{i=1}^{3\tilde{n}/2} C^{\tilde{n}^{3\tilde{n}/2-i} b'_i} \equiv \prod_{i=1}^{3\tilde{n}/2} (C^{\tilde{n}^{3\tilde{n}/2-i}})^{b'_i} (\% \ M)$$

when $C_1 = … = C_{3\tilde{n}/2} = C$. Notice that due to $b'_i \leq \tilde{n}/2$, there must be $b'_i < \tilde{n}$.

Let $\bar{G}_1 \equiv \prod_{i=1}^{3\tilde{n}/2} C_i^{b'_i}$ (% $M$) be of the SPP, where $b'_1…b'_{3\tilde{n}/2}$ corresponds to $B_1…B_{\tilde{n}/2}$.

Because $\bar{G}_1 \equiv \prod_{i=1}^{3\tilde{n}/2} C_i^{b'_i}$ (% $M$) and $\bar{G} \equiv \prod_{i=1}^{3\tilde{n}/2} C_i^{b'_i}$ (% $M$) with $0 \leq b'_i \leq b'_i$ have the same structure, by calling $\bar{O}_a(\bar{G}_1, C_1, …, C_{3\tilde{n}/2}, M)$, $b'_1…b'_{3\tilde{n}/2}$ can be found.

According to Definition 9, there is $\hat{H}(\bar{G}_1 \equiv \prod_{i=1}^{3\tilde{n}/2} C_i^{b'_i} (\% \ M)) \leq_T^P \hat{H}(\bar{G} \equiv \prod_{i=1}^{3\tilde{n}/2} C_i^{b'_i} (\% \ M))$.

Further by transitivity, there is

$$\hat{H}(y \equiv g^x (\% \ M)) \leq_T^P \hat{H}(\bar{G} \equiv \prod_{i=1}^{\tilde{n}/2} (C_{3(i-1)+B_i})^{B_i} (\% \ M)),$$

namely the ASPP $\bar{G} \equiv \prod_{i=1}^{\tilde{n}/2} (C_{3(i-1)+B_i})^{B_i}$ (% $M$) is at least equivalent to the DLP in the same prime field in computational complexity. □

- The proof of Property 9.

*Proof:*

Let $r_1…r_{\tilde{n}/2} = 1…1$, then the ASPP

$$\bar{G} \equiv \prod_{i=1}^{\tilde{n}/2} (C_{r_i(3(i-1)+B_i) + \neg r_i(3(i-B_i)+B_i)})^{B_i} (\% \ M)$$

is reduced to the ASPP $\bar{G} \equiv \prod_{i=1}^{\tilde{n}/2} (C_{3(i-1)+B_i})^{B_i}$ (% $M$).

By Property 8 and the transitivity, there exists

$$\hat{H}(y \equiv g^x (\% \ M)) \leq_T^P \hat{H}(\bar{G} \equiv \prod_{i=1}^{\tilde{n}/2} (C_{r_i(3(i-1)+B_i) + \neg r_i(3(i-B_i)+B_i)})^{B_i} (\% \ M)),$$

namely the ASPP is at least equivalent to the DLP in the same prime field in complexity. □

### 6.2 Resisting LLL Lattice Basis Reduction

We know that after a lattice basis is reduced through the LLL algorithm, the final reduced base will contain the shortest or approximately shortest vectors, but among them does not necessarily exist the original solution to a subset sum problem because only if

① the solution vector for the SSP is the shortest, and
② the shortest vector is unique in the lattice,

will the original solution vector appear in the reduced base with large probability.

In the new cryptoscheme, there are $\tilde{n}$ = 96, 112, or 128 and $\lceil \lg M \rceil$ = 464, 544, or 640. Under the circumstances, the DLP and IFP can be solved in tolerable subexponential time, namely the DLP and IFP cannot resist the attack of adversaries.

We first consider the ASPP $\bar{G} \equiv \prod_{i=1}^{\tilde{n}/2} (C_{3(i-1)+B_i})^{B_i}$ (% $M$).

For convenience, extend $B_1 \ldots B_{\tilde{n}/2}$ to $b'_1 \ldots b'_{3\tilde{n}/2}$ by the following rule for $i = 1, \ldots, \tilde{n}/2$:

① when $B_i = 0$, let $b'_{3(i-1)+1} = b'_{3(i-1)+2} = b'_{3(i-1)+3} = 0$;
② when $B_i \neq 0$, let $b'_{3(i-1)+1} = b'_{3(i-1)+2} = b'_{3(i-1)+3} = 0$, and $b'_{3(i-1)+B_i} = B_i$.

In this way, the ASPP $\bar{G} \equiv \prod_{i=1}^{\tilde{n}/2} (C_{3(i-1)+B_i})^{B_i}$ (% $M$) is converted into

$$\bar{G} \equiv \prod_{i=1}^{3\tilde{n}/2} C_i^{b'_i} \ (\%\ M).$$

Let $g$ be a generator of the group $(\mathbb{Z}_M^*, \cdot)$.

Let

$$C_1 \equiv g^{u_1} \ (\%\ M), \ldots, C_{3\tilde{n}/2} \equiv g^{u_{3\tilde{n}/2}} \ (\%\ M), \text{ and } \bar{G} \equiv g^v \ (\%\ M).$$

Then, through a conversion in subexponential time, seeking $B_1 \ldots B_{\tilde{n}/2}$ from $\bar{G}$ is equivalent to seeking $b'_1 \ldots b'_{3\tilde{n}/2}$ from the congruence

$$u_1 b'_1 + \ldots + u_{3\tilde{n}/2} b'_{3\tilde{n}/2} \equiv v \ (\%\ \overline{M}), \tag{3}$$

where $v$ may be substituted with $v + k \overline{M}$ along with $k \in [0, 3\tilde{n}/2]$ [5].

Similar to Section 1, $\{u_1, \ldots, u_{3\tilde{n}/2}\}$ is called a compact sequence due to every $b'_i \in [0, \tilde{n}/4 + 1]$ [6], and solving Equation (3) for $b'_1 \ldots b'_{3\tilde{n}/2}$ is called an ASSP [3].

May also convert this ASSP into a SSP through splitting $u_i$ into bits, and thus according to $b'_i \in [0, \tilde{n}/4 + 1]$, the density of the related ASSP knapsack is defined as $D = \sum_{i=1}^{3\tilde{n}/2} \lceil \lg(\tilde{n}/4 + 1) \rceil / \lceil \lg M \rceil = (3\tilde{n}/2) \lceil \lg(\tilde{n}/4 + 1) \rceil / \lceil \lg M \rceil$. Namely,

$$D = 3\tilde{n} \lceil \lg(\tilde{n}/4 + 1) \rceil / (2 \lceil \lg M \rceil). \tag{4}$$

which is slightly different from Formula (2).

Concretely speaking, in the new cryptoscheme, there are
$D = 144 \times 5 / 464 \approx 1.5517 > 1$ for $\tilde{n} = 96$ and $\lceil \lg M \rceil = 464$;
$D = 168 \times 5 / 544 \approx 1.5441 > 1$ for $\tilde{n} = 112$ and $\lceil \lg M \rceil = 544$;
$D = 192 \times 6 / 640 \approx 1.8000 > 1$ for $\tilde{n} = 128$ and $\lceil \lg M \rceil = 640$.

Therefore, Equation (3) does represent an ASSP of high density, which indicates that many different subsets will have the same sum, and probability that the original solution vector will occur in the final reduced lattice basis is nearly zeroth. Meanwhile, our experiment demonstrates that the original solution vector does not occur in the final reduced base [22].

Because $\bar{G} \equiv \prod_{i=1}^{\tilde{n}/2} (C_{3(i-1)+B_i})^{B_i}$ (% $M$) is only a special case of the ciphertext

$$\bar{G} \equiv \prod_{i=1}^{\tilde{n}/2} (C_{r_i(3(i-1)+B_i) + \neg r_i(3(i - B_i) + B_i)})^{B_i} \ (\%\ M),$$

the latter is also able to resist the LLL lattice basis reduction.

### 6.3 Avoiding Meet-in-the-middle Attack

Meet-in-the-middle dichotomy was first developed in 1977 [23]. Section 3.10 of [14] puts forward a meet-in-the-middle attack on a subset sum problem. It is not difficult to understand that the time complexity of the above algorithm is $O(n 2^{n/2})$.

Likewise, currently the versatile meet-in-the-middle dichotomy may be utilized to assault the ASSP $\bar{G} \equiv \prod_{i=1}^{\tilde{n}/2} (C_{r_i(3(i-1)+B_i) + \neg r_i(3(i - B_i) + B_i)})^{B_i}$ (% $M$) with entries

$$(\prod_{i=1}^{\tilde{n}/4} (C_{r_i(3(i-1)+B_i) + \neg r_i(3(i - B_i) + B_i)})^{B_i}, (r_1, \ldots, r_{\tilde{n}/4}), (B_1, \ldots, B_{\tilde{n}/4}))$$

for $(r_1, \ldots, r_{\tilde{n}/4}) \in \{0, 1\}^{\tilde{n}/4}$ and $(B_1, \ldots, B_{\tilde{n}/4}) \in \{00, 01, 10, 11\}^{\tilde{n}/4}$ when $B_{\tilde{n}/4} \neq 00$ and $B_{\tilde{n}/2} \neq 00$ which occurs with probability $9/16 = 0.5625$. Obviously, the random bit string $r_1\ldots r_{\tilde{n}/4}$ extends the scope of exhaustive search. Further, It is easy to see that the running time of this attack task is $O(\tilde{n}2^{\tilde{n}/2}2^{\tilde{n}/4}\lg^2 M) = O(\tilde{n}2^{3\tilde{n}/4}\lg^2 M)$ bit operations.

Concretely speaking,

when $\tilde{n} = 96$ namely $n = 80$ with $\lceil \lg M \rceil = 464$, $T_m = 2^7 2^{3 \times 96/4}(2^9)^2 = 2^{97} > 2^{80}$ bos;

when $\tilde{n} = 112$ namely $n = 96$ with $\lceil \lg M \rceil = 544$, $T_m = 2^7 2^{3 \times 112/4}(2^{10})^2 = 2^{111} > 2^{96}$ bos;

when $\tilde{n} = 128$ namely $n = 112$ with $\lceil \lg M \rceil = 640$, $T_m = 2^8 2^{3 \times 128/4}(2^{10})^2 = 2^{124} > 2^{112}$ bos.

Therefore, the new cryptoscheme can resist the meet-in-the-middle attack.

## 6.4 Avoiding Adaptive-chosen-ciphertext Attack

Most of public key cryptoschemes may probably be faced with adaptive-chosen-ciphertext attack [24]. However, It is lucky the Cramer-Shoup asymmetric encryption scheme is very indistinguishable and nonmalleable [25], and proven to be secure against the adaptive-chosen-ciphertext attack under the cryptographic assumptions [26]. So is the OAEP+ scheme [27].

### 6.4.1 Indistinguishability of Ciphertexts

In the encryption process of a JUOAN plaintext, ① a random padding string of $\tilde{n} - n$ bits is appended to the terminal of the JUOAN plaintext, which changes the original plaintext to an extended plaintext, and ② a random permutation string of $\tilde{n}/2$ bits is introduced into the arrangement of bit-pairs of the extended plaintext, which is equivalent to the thing that the order of triple items of a public key is varied along with every encryption.

Due to the interlacement of 00-pairs and non-00-pairs and the randomicity of bit string generation, the padding string and the permutation string make one identical original plaintext be able to correspond to at most $2^{\tilde{n}/4}2^{\tilde{n}-n}$ (exponential in $n$) different ciphertexts. It will take the running time of $O(\tilde{n}2^{\tilde{n}/2}2^{\tilde{n}-n}\lg^2 M)$ bit operations exhaustively to search all the possible ciphertexts of an original plaintext. Therefore, the correspondence between any arbitrary ciphertext and a related original plaintext are indistinguishable in subexponential time.

Concretely speaking, the running time of searching all the ciphertexts of an original plaintext is

$T_s = (96)2^{96/2}2^{96-80}(464)^2 \approx 2^{88} > 2^{80}$ for $n = 80$, $\tilde{n} = 96$, and $\lceil \lg M \rceil = 464$;

$T_s = (112)2^{112/2}2^{112-96}(544)^2 \approx 2^{98} > 2^{96}$ for $n = 96$, $\tilde{n} = 112$, and $\lceil \lg M \rceil = 544$;

$T_s = (128)2^{128/2}2^{128-112}(644)^2 \approx 2^{108} \approx 2^{112}$ for $n = 112$, $\tilde{n} = 128$, and $\lceil \lg M \rceil = 640$.

### 6.4.2 Nonmalleability of Ciphertexts

An encryption scheme is said to be malleable if it is possible for an adversary to transform a ciphertext into another ciphertext revertible to a related plaintext. That is, given an cipher-text of a plaintext $\underline{m}$, it is possible to generate another ciphertext which can decrypt to the plaintext $f(\underline{m})$ without necessarily knowing or learning $\underline{m}$, where $f$ is a known function [25].

By way of examples, let a RSA ciphertext $\bar{C} = \underline{m}^e \% N$, then
$$z^e \bar{C} = (z\underline{m})^e \% N$$
is a malleation of $\bar{C}$, which decrypts to $f(\underline{m}) = z\underline{m} \% N$.

Again let an ElGamal ciphertext $\bar{C} = (g^r, \underline{m}y^r \% p)$, then
$$z\bar{C} = (g^r, z\underline{m}y^r \% p)$$
is a malleation of $\bar{C} = (g^r, \underline{m}y^r \% p)$, which decrypts to $f(\underline{m}) = z\underline{m} \% p$. Thus, if $\underline{m}' = z\underline{m} \% p$ is known, then $\underline{m} = \underline{m}'z^{-1} \% p$ is found.

In the new scheme, there is the ciphertext
$$\bar{G} = E(\dot{B}) = \prod_{i=1}^{\tilde{n}/2} (C_{r_i(3(i-1)+B_i) + \neg r_i(3(i-B_i)+B_i)})^{B_i} \% M$$
which takes a bit-pair as an operation unit, and where $\dot{B} = B_1\ldots B_{\tilde{n}/2}$ is a related extended plaintext.

Thus, evidently the plaintext function $f(\dot{B}) = z\dot{B} \% M$ is not suitable for $\bar{G}$. Again considering that $B_j$ occurs in the subscript of multiplied $C_i$, and moreover is relevant to the random bit string $r_1\ldots r_{\tilde{n}/2}$ that can be guessed only in exponential time, it is impossi-ble to exist other plaintext function $f(\dot{B})$ which corresponds to the malleation of $E(\dot{B}) = \bar{G}$.

### 6.4.3 Proof of the Semantical Security

If the security requirement of a cryptoscheme can be stated formally in an antagonistic model, as opposed to heuristically, with clear assumptions that certain computational problems are intractable, and an adversary has access to he algorithms of the cryptoscheme as well as enough computational resources, the cryptoscheme possesses provable security [28][29].

***Definition 11***: A cryptoscheme is said to be semantically secure if an adversary who knows the encryption algorithm of the cryptoscheme and is in possession of a ciphertext is unable to determine any information about the related plaintext [28].

It is subsequently demonstrated that semantic security is equivalent to another definition of security called ciphertext indistinguishability [30]. If a cryptoscheme has the property of indistinguishability, then an adversary will be unable to distinguish a pair of ciphertexts based on the two plaintexts encrypted by a challenger.

A chosen plaintext attack (CPA) is an attack model for cryptanalysis which presumes that the attacker has the capability to choose arbitrary plaintexts to be encrypted and obtain the corresponding ciphertexts that are expected to decrease the security of an encryption scheme [29].

***Definition 12***: A cryptoscheme is said to be IND-CPA (indistinguishable under chosen plaintext attack), namely semantically secure against chosen plaintext attack, if the adversary cannot determine which of the two plaintexts was chosen by a challenger, with probability significantly greater than 1/2, where 1/2 means the success rate of random guessing [29][31].

For a probabilistic asymmetric cryptoscheme based on computational security, indistinguishability under chosen plaintext attack is illuminated by a game between an adversary and a challenger, where the adversary is regarded as a probabilistic polynomial time Turing machine, which means that it must complete the game and output a guess within a polynomial number of operation steps.

Notice that for the JUOAN cryptoscheme, the adversary may be also regarded as a probabilistic subexponential time Turing machine since no subexponential time solution to the MPP or ASPP is found so far.

***Theorem 1***: The JUOAN cryptoscheme is semantically secure against chosen plaintext attack on the assumption that the MPP and ASPP cannot be solved in subexponential time.

*Proof*:

Let $E(k_p, m)$ represents the encryption of a message (plaintext) $m$ under the public key $k_p$.

A game between an adversary and a challenger is given as follows.

① The challenger calls the key generation algorithm with the parameters $n$, $\tilde{t}$, and $Б$, obtains a key pair $(k_p, k_s)$, publishes $k_p = (\{C_1, …, C_{3\tilde{n}/2}\}, M)$ to the adversary, and retains $k_s$ for himself.

② The adversary may perform any number of encryptions or other compatible operations.

③ Eventually, the adversary chooses any two distinct $n$-bit plaintexts ($m_0$, $m_1$), and submits them to the challenger.

④ The challenger selects a bit $x \in \{0, 1\}$ uniformly at random, and sends the challenge ciphertext

$$\bar{G} = E(k_p, m_x) = \prod_{i=1}^{\tilde{n}/2} (C_{r_i(3(i-1)+B_i) + \neg r_i(3(i-B_i)+B_i)})^{B_i} \% M$$

back to the adversary, where $m_x = b_1…b_n$.

⑤ The adversary is free to perform any number of additional computations or encryptions. Finally, it outputs a guess for the value of $x$. Therefore need to analyze the probability of hitting $x$.

Because the intractabilities MPP and ASPP have no subexponential time solutions, neither can the adversary decrypt $\bar{G}$ for $m_x$ with a private key, nor can directly solve

$$\bar{G} \equiv \prod_{i=1}^{\tilde{n}/2} (C_{r_i(3(i-1)+B_i) + \neg r_i(3(i-B_i)+B_i)})^{B_i} (\% M)$$

for $m_x$ ($= b_1…b_n = B_1…B_{n/2}$).

It is known from the encryption algorithm that one identical plaintext may correspond to at most $2^{\tilde{n}/4} 2^{\tilde{n}-n}$ different ciphertexts, where $\tilde{n} = n + 16$, and it will need the running time of $O(\tilde{n} 2^{\tilde{n}/2} 2^{\tilde{n}-n} \lg^2 M)$ bit operations to verify all the possible ciphertexts of a plaintext. Thus the probability that the adversary hits $x$ with guessing is only $(1 / 2) + (1 / 2^{\tilde{n}/4} 2^{\tilde{n}-n})$, where $2^{\tilde{n}/4} 2^{\tilde{n}-n}$ is exponential in $n$, which means that $1/2^{\tilde{n}/4} 2^{\tilde{n}-n}$ is a negligible function of $n$, and for every (nonzero) polynomial function $poly(n)$ (notice that in the JUOAN cryptoscheme, it may be also a subexponential function), there exists $n_0$ such that $1/2^{\tilde{n}/4} 2^{\tilde{n}-n} < 1 / poly(n)$ for all $n > n_0$.

In summary, the JUOAN public key cryptoscheme is semantically secure, namely IND-CPA.  □

# 7 Conclusion

The new cryptoscheme builds its security firmly on two intractabilities:

① the MPP $C_i = (A_i W^{\ell(i)})^\delta \% M$ with $A_i \in \Lambda$ and $\ell(i)$ from $\Omega$, and

② ASPP $\bar{G} \equiv \prod_{i=1}^{\tilde{n}/2} (C_{r_i(3(i-1)+B_i)+\neg r_i(3(i-B_i)+B_i)})^{\mathcal{B}_i} (\% M)$.

No subexponential time solutions to them are found, and there exist only exponential time solutions so far [32].

The new cryptoscheme utilizes a bit-pair string to decrease the bit-length of the modulus $M$, exploits a bit-pair shadow string to guard against the LLL lattice basis reduction attack, and adopts the approaches of introducing a random bit string into an encryption and appending a random bit string to a plaintext to avoiding the adaptive-chosen-ciphertext attack and the meet-in-the-middle dichotomy.

As $\tilde{n}$ = 96, 112, or 128, there exists $\lceil \lg M \rceil$ = 464, 544, or 640, which assures that when a JUOAN ciphertext $\bar{G}$ with $r_1 \ldots r_{\tilde{n}/2}$ = 1…1 is converted into an ASSP through a discrete logarithm, the density of a related ASSP knapsack is pretty high, and larger than 1.

There always exists contradiction between time and security, so does between space and security, and so does between time and space. We attempt to find a balance which is none other than a delicate thing among time, space, and security.

## Acknowledgment

The authors would like to thank the Academicians Jiren Cai, Zhongyi Zhou, Jianhua Zheng, Zhengyao Wei, Changxiang Shen, Binxing Fang, Guangnan Ni, Andrew C. Yao, Xicheng Lu, Wen Gao, Jinpeng Huai, Wenhua Ding, and Xiangke Liao for their important advice and helps.

The authors also would like to thank the Professors Dingyi Pei, Jie Wang, Ronald L. Rivest, Moti Yung, Adi Shamir, Dingzhu Du, Mulan Liu, Huanguo Zhang, Yixian Yang, Hanliang Xu, Dengguo Feng, Xuejia Lai, Yongfei Han, Yupu Hu, Dongdai Lin, Ping Luo, Jianfeng Ma, Rongquan Feng, Lusheng Chen, Chuankun Wu, Lin You, Wenbao Han, Bogang Lin, Lequan Min, Qibin Zhai, Hong Zhu, Renji Tao, Zhiying Wang, Quanyuan Wu, and Zhichang Qi for their important suggestions and corrections.